\PassOptionsToPackage{unicode}{hyperref}
\PassOptionsToPackage{hyphens}{url}
\documentclass[
  12pt,
  a4paper]{article}
\usepackage{amsmath,amssymb}
\usepackage{iftex}
  \usepackage[T1]{fontenc}
  \usepackage[utf8]{inputenc}
    \usepackage{textcomp} 
    \usepackage{newunicodechar}
\usepackage{lmodern}
\ifPDFTeX\else
\fi
\IfFileExists{upquote.sty}{\usepackage{upquote}}{}
\IfFileExists{microtype.sty}{
  \usepackage[]{microtype}
  \UseMicrotypeSet[protrusion]{basicmath} 
}{}
\makeatletter
\@ifundefined{KOMAClassName}{
  \IfFileExists{parskip.sty}{%
    \usepackage{parskip}
  }{
    \setlength{\parindent}{0pt}
    \setlength{\parskip}{6pt plus 2pt minus 1pt}}
}{
  \KOMAoptions{parskip=half}}
\makeatother
\usepackage{xcolor}
\usepackage[left=3cm,right=3cm,top=2cm,bottom=2cm]{geometry}
\usepackage{longtable,booktabs,array}
\usepackage{calc} 
\usepackage{etoolbox}
\makeatletter
\patchcmd\longtable{\par}{\if@noskipsec\mbox{}\fi\par}{}{}
\makeatother
\IfFileExists{footnotehyper.sty}{\usepackage{footnotehyper}}{\usepackage{footnote}}
\makesavenoteenv{longtable}
\usepackage{graphicx}
\makeatletter
\newsavebox\pandoc@box
\newcommand*\pandocbounded[1]{
  \sbox\pandoc@box{#1}%
  \Gscale@div\@tempa{\textheight}{\dimexpr\ht\pandoc@box+\dp\pandoc@box\relax}%
  \Gscale@div\@tempb{\linewidth}{\wd\pandoc@box}%
  \ifdim\@tempb\p@<\@tempa\p@\let\@tempa\@tempb\fi
  \ifdim\@tempa\p@<\p@\scalebox{\@tempa}{\usebox\pandoc@box}%
  \else\usebox{\pandoc@box}%
  \fi%
}
\def\fps@figure{htbp}
\makeatother
\NewDocumentCommand\citeproctext{}{}

\makeatletter
 \let\@cite@ofmt\@firstofone
 \def\@biblabel#1{}
 \def\@cite#1#2{{#1\if@tempswa , #2\fi}}
\makeatother
\newlength{\cslhangindent}
\newlength{\csllabelwidth}
\newenvironment{CSLReferences}[2] 
 {\begin{list}{}{%
  \setlength{\itemindent}{0pt}
  \setlength{\leftmargin}{0pt}
  \setlength{\parsep}{0pt}
  \ifodd #1
   \setlength{\leftmargin}{\cslhangindent}
   \setlength{\itemindent}{-1\cslhangindent}
  \fi
  \setlength{\itemsep}{#2\baselineskip}}}
 {\end{list}}
\usepackage{calc}

\usepackage[T1]{fontenc}
\usepackage[utf8]{inputenc}
\usepackage{lmodern}
\usepackage{ifxetex}
\ifxetex
  \usepackage{fontspec}
\else
  \usepackage[T1]{fontenc}
  \usepackage[utf8]{inputenc}
  \usepackage{lmodern}
  \usepackage{textgreek}
\fi
\DeclareUnicodeCharacter{0301}{\'{}} 
\usepackage{float}
\usepackage{bookmark}
\IfFileExists{xurl.sty}{\usepackage{xurl}}{} 
\hypersetup{
  hidelinks,
  pdfcreator={LaTeX via pandoc}}

\author{}
\date{\vspace{-2.5em}}

\begin{document}

\section*{Relationship Between Leisure Activities, Stress Management Methods, Study Methods, and Methods of Learning New Things Among First-Year Statistics Students}\label{relationship-between-leisure-activities-stress-management-methods-study-methods-and-methods-of-learning-new-things-among-first-year-statistics-students}
\addcontentsline{toc}{section}{Relationship Between Leisure Activities, Stress Management Methods, Study Methods, and Methods of Learning New Things Among First-Year Statistics Students}

Thiyanga S. Talagala

Department of Statistics, Faculty of Applied Sciences, University of Sri Jayewardenepura, Sri Lanka

Corresponding author address:\newline Department of Statistics, Department of Statistics, Faculty of Applied Sciences, University of Sri Jayewardenepura, Gangodawila, Nugegoda, CO 10250, Sri Lanka

Corresponding author email address:\newline \href{mailto:ttalagala@sjp.ac.lk}{\nolinkurl{ttalagala@sjp.ac.lk}}

\newpage

\section*{Abstract}\label{abstract}
\addcontentsline{toc}{section}{Abstract}

The interplay between leisure activities, stress management methods, studying methods, and methods of learning new things is crucial and affects performance in all aspects of life. On the other hand, data science and statistics are rapidly growing fields with high demands across universities. Thus, this study aimed to identify the similarities and dissimilarities between the four dimensions: leisure activities, stress management methods, studying methods and methods of learning new things. The participants of this study were first-year undergraduates studying statistics at one of the universities in Sri Lanka. There were 117 students in the sample (female-65, male-52). A self-reported questionnaire was used to collect data. First, individual responses for each question under each dimension were visualized using tile maps separately for males and females to identify similarities and dissimilarities in responses. Next, individuals were clustered based on the responses for each dimension separately. Finally, all resulting clusters were re-clustered to identify the relationships between the dimensions. In all cluster analyses, we used Jaccard distance with
hierarchical clustering using the complete linkage method. The results were visualized using tile maps. Across all four dimensions we considered, the top activities were either listening to
music or lectures and watching videos or TV shows, suggesting that individuals are
introverts and passive learners. There was no strong relationship between these dimensions. By identifying these clusters and relationships, educators can tailor instructional approaches to enhance engagement and effectiveness in diverse learning environments.

\section*{Keywords}\label{keywords}
\addcontentsline{toc}{section}{Keywords}

Teaching, Learning, Statistics, Visualization, Clustering

\newpage

\section*{Highlights}\label{highlights}
\addcontentsline{toc}{section}{Highlights}

\begin{itemize}
\item
  Clustering revealed distinct leisure profiles containing both introverts and extroverts.
\item
  Engagement in kinesthetic activities were low.
\item
  Clustering and tile map visualizations were used to visualize individual differences.
\item
  No clear relationship between leisure, stress management, study techniques, and learning methods.
\end{itemize}

\section*{Graphical abstract}\label{graphical-abstract}
\addcontentsline{toc}{section}{Graphical abstract}

\pandocbounded{\includegraphics[keepaspectratio]{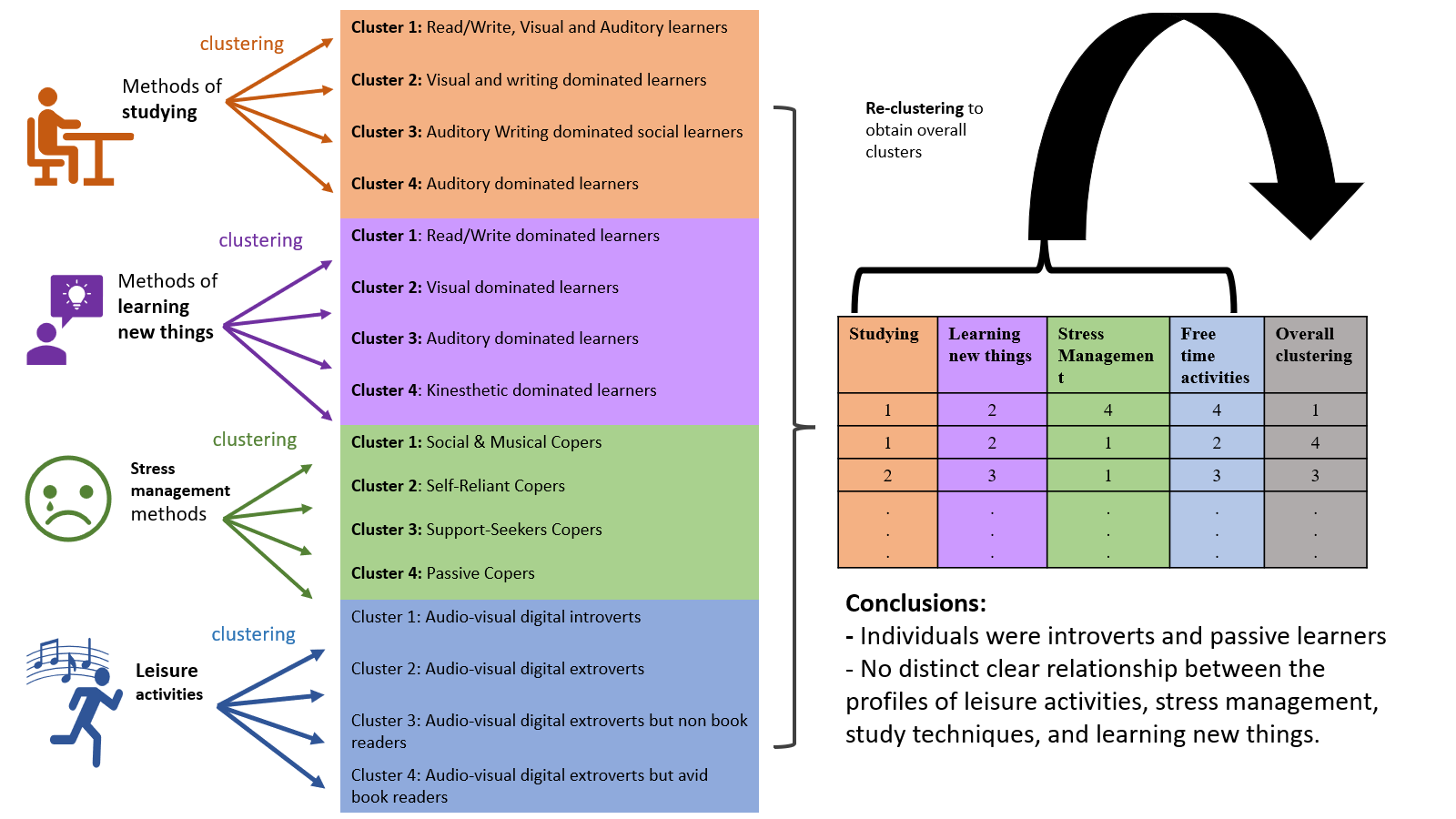}}

\newpage

\section{Introduction}\label{intro}

Having a balanced lifestyle is crucial for one's well-being. Interplay between learning styles, leisure activities, and stress management is crucial for having balance and quality lifestyles. Although there are several studies done in other disciplines about studying methods, learning styles, leisure activities, and stress management methods, there has been limited research done in the field of statistics even though this is an important field providing the base to the demanding fields such as data science, artificial intelligence, etc. To fill this gap, this study explores four important dimensions of students life activities. They are i) leisure activities, ii) stress management methods, iii) studying methods, and iv) methods of learning new things among first-year statistics students who have just enrolled at the university. Furthermore, we explore the relationships between these dimensions. Here, ``methods of learning new things'' is referred to as learning something for the first time, while studying is referred to as ``working to understand and remember it well afterward.'' Findings can inform university policies and strategies aimed at improving the overall student experience from orientation programs to ongoing support services.

\subsection{Leisure activities}\label{leisure-activities}

Fosnacht, McCormick, and Lerma (2018) stated that ``students' time expenditures influence their learning and development''. Caldwell, Smith, and Weissinger (1992) conducted a study to explore the relationship between leisure activities and perceived health of college students. The results showed participation in specific types of leisure activities among college students associated with higher levels of physical, mental, and social health. A study conducted at a large research university in the Midwest at five-year intervals beginning in 1971 by Hendel and Harrold (2004) found that there was a significant decrease in reading-related leisure activities among college students over the last three decades, while electronically based activities saw the most significant increases. Lehto et al. (2014) conducted a study to find the relationship between anticipated leisure benefits, student life stress, and leisure activity participation. Their results showed there was no direct relationship between student stress and their leisure participation. Kuo and Tang (2014) conducted a study on possible relationships among personality traits, Facebook usages, and leisure activities. They found there was a dependency between leisure activities and personality traits. An extensive review done with 73,304 articles were retrieved and 64 articles representing 47 countries by Hulteen et al. (2017) reported that the child and adolescent leisure activities were highly dependent upon region. Kim and Brown (2018) conducted a study to identify the associations between leisure, stress, and health behavior among university students at a Midwestern U.S. university campus. The study suggested when people find their leisure activities enjoyable and fulfilling, they may experience lower stress levels, which contributes to better overall health. Chan et al. (2023) conducted a study to explore the associations between leisure activities and substance use over time, examining whether these associations vary based on residential status (i.e., living with parents vs.~independently). Four leisure categories were included in the study: (1) socializing with friends, (2) spending time with family, (3) sports, and (4) non-sports. Terzi et al. (2024) explored the interplay between free time management, leisure satisfaction, and quality of life by gender,
age, and participation in physical activity, and the results showed significant relationships between gender, free time management, and life quality.

\subsection{Stress management methods}\label{stress-management-methods}

Stress is a persistent serious health problem among university students (Bistricky et al. 2018). Mosley Jr et al. (1994) examined stress, coping, depression, and somatic distress among 69 third year medical students at the University of Mississippi School of Medicine. The coping was assessed using eight primary factors in which four are classified as engagement strategies (i. cognitive-restructuring: manage stress altering the meaning, ii. problem-solving: direct attempt to eliminate the source o stress by changing the situation, iii. social support, iv. express-emotion) and others are classified as disengagement strategies (i. wishful-thinking, ii. problem-avoidance: behavioral or cognitive avoidance, iii. self-criticism, iv. social withdrawal: avoidance of others). They found coping strategies classified as engagement strategies were linked to lower levels of depressive symptoms. After analyzing 100 assessments of coping, Skinner et al. (2003) constructed a categorization that distinguishes coping strategies, providing a framework for understanding how individuals manage stress. Ramos (2011) explored the differences in perceived stress and coping styles among non-traditional graduate students in both on-campus and distance-learning programs. The results showed there was no significant difference between the coping styles and the perceived stress levels of non-traditional graduate students who enrolled in distance-learning and on-campus programs. Sikander and Aziz (2012) highlighted that one of the most common coping strategies used by the nursing students was discussing feelings with friends or classmates. A study done on third-year teacher education students at the University of Ghana identified praying or meditating and self-diverting actions as the most commonly used coping strategies (Amponsah et al. 2020). Adasi et al. (2020) explored the coping strategies employed by teachers. The author classified coping strategies into 11 categories such as: i. Active coping: Doing something about the situation, making a move
to discredit stressor,
ii. Positive reframing: Observing something great in what is happening, learning from experience, iii.
Substance abuse Using tobacco/alcohol/drug to feel better, iv. Humour Making fun of the situation, v.
Given up coping: Giving up the attempt to do anything about the
situation, vi. Emotional support: Getting emotional support/ advice from family and
friends, vii. Instrumental support: Getting help and exhortation from academic staff,
viii. Self-distraction Doing something to take my mind off the
circumstance, ix. Religion: Praying and meditating, x.
Venting Expressing negative feelings: indicating outrage at
things/people, xi. Accepting the circumstance as it is, learning to live
with it. Among them, ``Positive reframing'' and ``Religion'' were the most commonly used strategies. Waterhouse and Samra (2024) conducted a scope review to explore ``what is known about how university students cope with stress'' among nursing, medical and dentistry degrees. They summarized the coping strategies mapping to Skinner et al. (2003) categorization of coping. Waterhouse and Samra (2024) highlighted that the majority of research on coping focuses on the types of coping strategies, with limited exploration of what works for whom and in which specific domains of stress. Further, most of the studies explore the coping strategies among nursing, medical, and dentistry fields (Abdelhafez et al. 2021; Ahmad, Md Yusoff, and Razak 2011; Sikander and Aziz 2012; Al-Gamal, Alhosain, and Alsunaye 2018; Alshahrani, Cusack, and Rasmussen 2018). To address this gap, we explore coping strategies and their relation to other dimensions, such as leisure activities, study methods, and learning new things, of the undergraduates who are taking statistics as a main subject in the first two years of the degree program.

\subsection{Learning and studying methods}\label{learning-and-studying-methods}

Different people learn things in different ways (Van Zwanenberg, Wilkinson, and Anderson 2000; Pashler et al. 2008; Yousef 2016). There are different categorizations and models of learning styles. Kolb (1976) introduced a four-stage experimental learning model. The four stages were: i) concrete experience - doing/having an experience, ii) reflective observation - reviewing/reflecting on the experience, iii) active experimentation - planning/trying out what you have learned and iv) abstract conceptualization - concluding/ learning from the experience. Honey and Mumford's clever theory (Honey and Mumford 1989) introduced 4 types of learner. They are, i) activist: having an experience, ii) reflector: reviewing the experience, iii) pragmatist: planning the next step and iv) theorist: concluding from the experience. Dunn and Dunn learning styles model (Dunn 1990) lists five key areas that influence how individuals learn: environmental, emotional, sociological, physiological, and psychological. Richard M. Felder (1993) presented five dimensions of learning styles as: i) sensing/intuitive,
ii) visual/verbal, iii) inductive/deductive, iv) active/reflective, and v) sequential/global. Further more, author highlighted the students preference and level of preference (strong, moderate, or almost nonexistent) could change over time. Most of the research in engineering filed used Felder-Silverman learning styles model to explore the learning styles of engineering students (Domı́nguez, Robles-Gómez, and Pastor-Vargas 2025; Katsioloudis and Fantz 2012). Fleming (2006) in his VARK model emphasized four main modes of learning styles visual, aural, reading, and kinesthetic. Many researchers used Neil Flemings's VARK model to identify students' learning styles (Prithishkumar and Michael 2014; Moayyeri 2015; Espinoza-Poves, Miranda-Vı́lchez, and Chafloque-Céspedes 2019; Klement 2014; Othman and Amiruddin 2010).

Jaju, Kwak, and Zinkhan (2002) used Hofstede's cross-cultural framework and Kolb's experiential learning model to investigate the cross-cultural Comparison between the US, India, and Korea of undergraduate business students. The results showed there is a difference in learning styles between the three countries. Richard M. Felder (2002) conducted a study to answer three question corresponds to the engineering students. The question were ``1) Which aspects of learning style
are particularly significant in engineering education?
2) Which learning styles are preferred by most students and which are
favored by the teaching styles of most
professors?
3) What can be done to reach students whose learning styles are not
addressed by standard methods of
engineering education?''. The results indicated that most engineering students were
visual (pictures, diagrams, graphs, demonstrations), sensing (sights, sounds, physical sensations,), inductive (facts and observations are given, underlying principles
are inferred), and active (through
engagement in physical activity or discussion). Naik (2013) investigated the influence of western, middle-eastern, and eastern cultures on learning style distribution of business students with respect to the Felder-Silverman learning style model (FSLSM)(R. M. Felder and Silverman, n.d.). The results showed that all three cultures had a balanced learning style in each of the four learning style dimensions of the FSLSM. Yousef (2016) explored the learning style preferences of statistics students at the UAE model based on FSLSM. The results showed that undergraduate statistics students at UAE University had balanced preferences across the four dimensions of learning styles. Additionally, the findings implied that there were no statistically significant variations among the four learning style dimensions with respect to demographic and academic characteristics, with the exception of the sensing-intuitive and active-reflective aspects related to the type of high school (private vs.~public). Furthermore, Yousef (2016) concluded there was no
statistically significant differences between male and female statistics students along the FSLSM four
dimensions of learning styles. Seemiller et al. (2019) conducted a comparative analysis to investigate how generation Z college students prefer to learn in the United States and Brazil. The authors explored characteristics, motivations, interpersonal styles, learning styles, and learning methods. Göğüş and Ertek (2020) introduced a scoring approach for the assessment of study skills and learning styles. Rohani et al. (2024) conducted a systematic review to identify learning preferences and strategies in the health data science discipline. The results showed that most health data science learners preferred visual presentations as their primary mode of learning. Identification of dominated learning styles can benefit teachers in differentiating their teaching method and help students in improving their results (Mašic, Polz, and Becirovic 2020).

\subsection{Gender differences}\label{gender-differences}

Wehrwein, Lujan, and DiCarlo (2007) conducted a study to identify gender differences in learning style preferences among undergraduate physiology students using the VARK questionnaire. The results showed a majority of male students are multi-modal learners, whereas a majority of female students preferred single-mode learners, and hence concluded that male and female students have significantly different learning styles. Agahi and Parker (2008) conducted a study to examine the association between participation in leisure activities and mortality risk among elderly men and women. The results showed gender differences in the association between leisure activities and mortality. Further, the results showed women benefited from social activities, whereas men seem to have benefited from solitary activities. Adasi et al. (2020) conducted a study to identify gender differences in stressors and coping strategies among teacher education students at the University of Ghana. They found, although, females had higher overall perceived stress levels than males, there was no significant difference between the coping strategies used by them. Idrizi, Filiposka, and Trajkovikj (2023) conducted a study to identify gender impact on STEM (science, technology, engineering and mathematics) online learning. The study also showed there was a gender differences in learning styles, where female students preferred the style of read/write while male students favored kinesthetic. Most of the studies explored the leisure activities of female and male adolescents (Athenstaedt, Mikula, and Bredt 2009; Finkel, Andel, and Pedersen 2018; Avital 2017).

\subsection{The present study}\label{the-present-study}

Through the literature review we have identified 5 main gaps in the field of the research.

\begin{enumerate}
\def\labelenumi{\arabic{enumi}.}
\item
  A handful of prior research investigate the individual differences in learning styles in the field of statistics even though the field of statistics has a high demand which also provides foundations to other fields such as Data Science, Artificial Intelligence, etc.
\item
  No attempt has been made to identify methods
  of learning new things, leisure activities, and
  stress coping styles among statistics students.
\item
  There is no research done on exploring the ``Methods of Learning New Things'' in STEM.
\item
  To the best of our knowledge, there has been no attempt to identify relationships between methods of studying, methods of learning new things, methods of managing stress and methods of spending free time across any disciplines.
\end{enumerate}

To fill the gap, through this study we try to answer the following research questions.

\begin{enumerate}
\def\labelenumi{\arabic{enumi}.}
\item
  What are the most commonly preferred \textbf{studying methods} among the statistics students, and are there any notable differences between male and female preferences?
\item
  What are the most commonly preferred \textbf{methods of learning new things} among the statistics students, and are there any notable differences between male and female preferences?
\item
  What are the most commonly preferred \textbf{leisure activities} among the statistics students, and are there any notable differences between male and female preferences?
\item
  What are the most commonly preferred \textbf{stress management methods} among the statistics students, and are there any notable differences between male and female preferences?
\item
  What are the \textbf{relationships between studying methods, methods of learning new things, leisure activities and stress management methods} ?
\end{enumerate}

The paper is organized as follows. The \autoref{intro} provides background of the study and relevant literature and objectives of the study. The \autoref{method}
details the data description and methodology employed in the study. The \autoref{results} presents the findings of the statistical analysis, followed by a discussion in \autoref{discussion}. Finally, \autoref{conclusions} offers conclusions and recommendations.

\section{Material and methods}\label{method}

\subsection{Participants}\label{participants}

The participants of this study were first-year undergraduates who are following statistics as a subject at one of the universities in Sri Lanka. All of these students followed Mathematics, Physics and Chemistry at advanced level examination.

The data collection measurement was a structured questionnaire in the year of 2024. The questionnaire was formatted as a Google form. Data were collected during the first four weeks of the semester. There were 130 students enrolled in total during that period. Among them 117 students completed the questionnaire. Among them 55.6\% (65) were female and 44.4\% (52) were male.

\subsection{Measures}\label{measures}

The questionnaire consists of both multiple response questions and single response questions. Our goal is to obtain a well rounded picture about the individual differences in studying, methods of learning new things, leisure activities, and stress management methods. Together, these are called the four dimensions of our study. Only one question is used to explore each dimension, avoiding repetitions and overlaps. All of the responses in response for each question can be categorized in to four categories. They are i) Visual: activities primarily involving seeing or visualizing, such as watching movies or TV shows or drawing; ii) Auditory: Activities involving listening or auditory engagement, like listening to music or podcasts; iii) Kinesthetic: Activities that involve physical movement or engagement, such as engaging in outdoor activities or sports. and iv) Read \& Write: Activities centered around reading, writing, or textual engagement, like reading books or spending time on social media platforms.

\subsubsection{Leisure activities}\label{leisure-activities-1}

To identify the leisure activities multiple response question ``How do you spend your free time?'' was used. This question was asked before inquiring about stress management methods, studying and methods of learning new things to reduce respondent drop-off at the very beginning of the questionnaire without making the respondents stressful.

\subsubsection{Stress management methods}\label{stress-management-methods-1}

To identify stress management methods multiple response question ``How do you usually manage stress?'' was used. The reason for using multiple-response questions is to reduce the non-response rate while obtaining a comprehensive understanding of stress management practices without feeling them anxious about providing answers.

\subsubsection{Studying methods}\label{studying-methods}

To identify different studying methods multiple response question ``Please indicate which methods you typically do when studying. Select all that apply.'' was used. Multiple response questions provides a broader understanding of their preferences and habits when studying.

\subsubsection{Methods of learning new things}\label{methods-of-learning-new-things}

To identify methods of learning new things the multiple response question ``How do you prefer to learn new things?'' was used. This focused on capturing how individuals approach and absorb completely new information for the first time.

\subsubsection{Demographic details}\label{demographic-details}

In addition to above, gender of the respondent was recorded. This is the only demographic information collected. We know these respondents were within the age range 20 - 25 years.

\subsection{Statistical analyses}\label{statistical-analyses}

\begin{figure}
\includegraphics[width=1\linewidth]{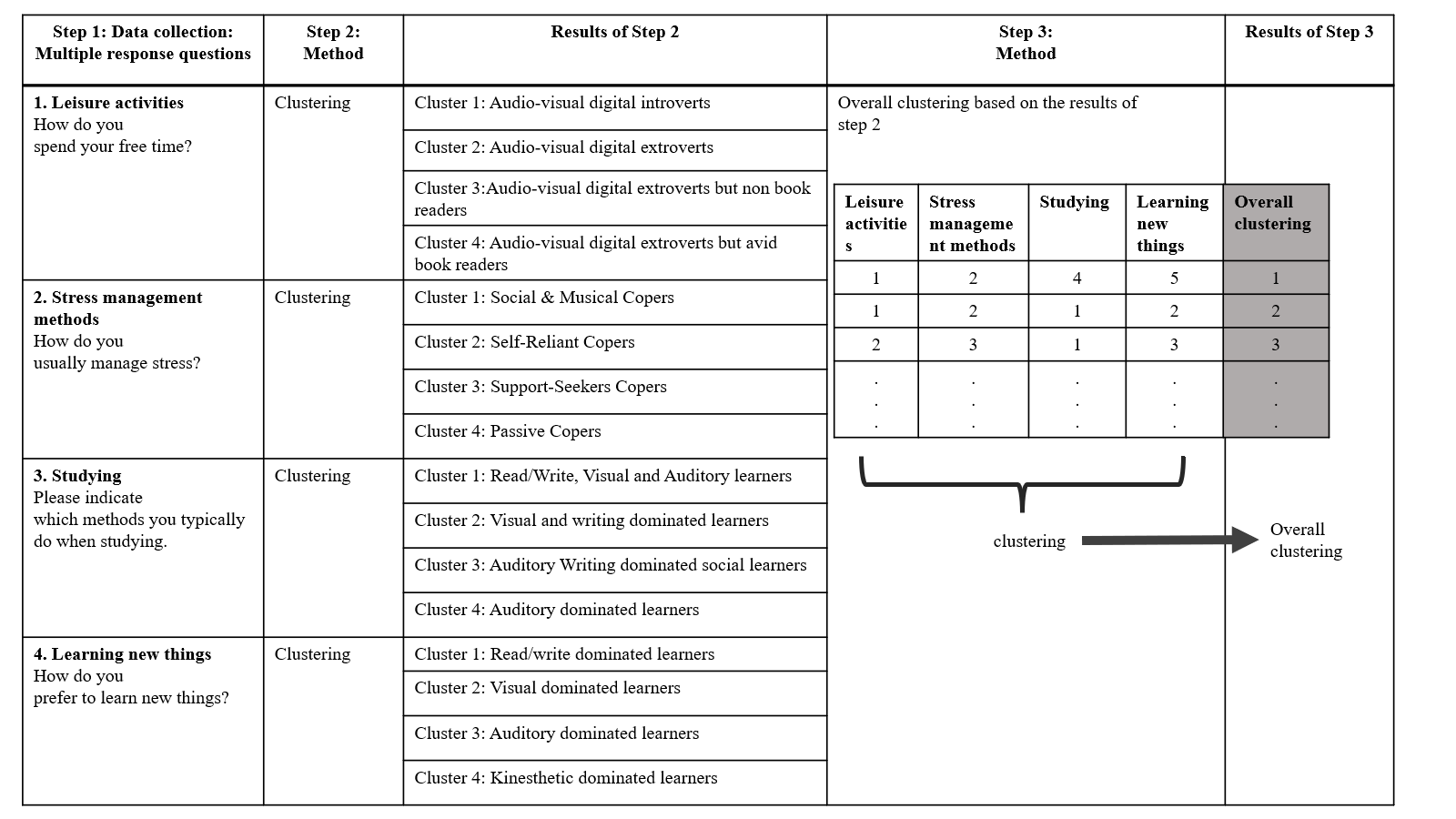} \caption{Diagram illustrating the application of cluster analysis: First, cluster analysis was performed separately on each individual dimension—studying, methods of learning new things, leisure activities, and stress management methods. Then, the resulting clusters from each dimension were further clustered together to identify overall similarities and dissimilarities among individuals.}\label{fig:methodology}
\end{figure}

First, individual responses for each question under each dimensions were visualized using tile maps separately for males and females to identify similarities and dissimilarities in responses. Next, individuals were clustered based on the responses for i) leisure activities, ii) stress management methods, iii) methods of learning new things and v) studying methods questions. Individuals were clustered four times according to each dimension. The reason for clustering individuals separately for each dimension---leisure activities, stress management methods, methods of learning new things, and studying methods--- helps uncover meaningful groups within each dimension without interference from unrelated dimensions. This helps to visualize different aspects of individual behavior, revealing patterns far deeper. Furthermore, this allows for a more focused and interpretable analysis of patterns within each dimension. If all responses were clustered together across categories, the results might be difficult to interpret, and important details could be missed out. Once separate clusters were identified, the resulting clusters from each dimension were further clustered together to identify overall similarities and dissimilarities among individuals. This helps to examine whether and how clusters from different categories overlap, revealing potential relationships between learning styles, stress management, and leisure activities. For example this helps to explore whether individuals belonging to a particular cluster in one dimension (e.g., learning styles) tend to fall into the same cluster in another dimension (e.g., stress management methods). This approach helps identify potential associations between learning styles, methods of learning new things, stress management techniques and leisure time activities. A tile map visualisation was used to present results. The cluster analysis procedure was demonstrated in Figure \ref{fig:methodology}. In all cluster analysis we use Jaccard distance with hierarchical clustering using the complete linkage method. All of the data analysis were performed using the R programming software (R Core Team (2024a)). The tidyr and dplyr in tidyverse package collection was used for data wrangling (Wickham et al. (2019)). The ggplot2 package (R Core Team (2024b)) was used for data visualisations. The R package cluster (Maechler et al. (2023)) was used for cluster analysis.

\section{Results}\label{results}

\subsection{Leisure activities}\label{leisure-activities-2}

\begin{figure}
\includegraphics[width=1\linewidth]{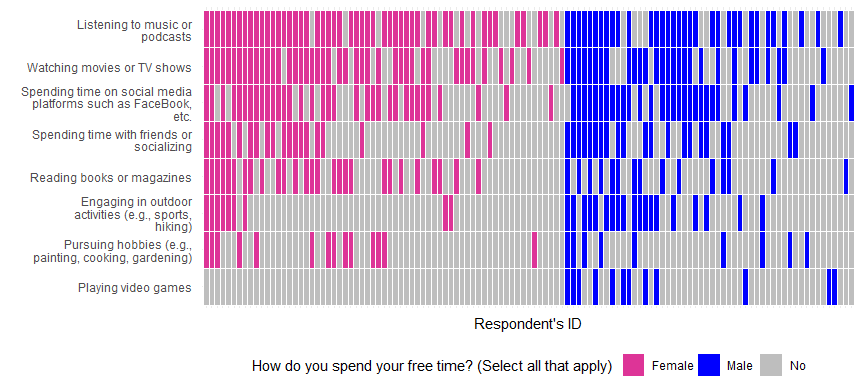} \caption{Heat map of responses to the multiple response question: How do you spend your free time? by gender. The Y-axis represents the responses, while the X-axis represents the respondent IDs. Most males and females listen to music and podcasts.}\label{fig:freetimeg}
\end{figure}

Figure \ref{fig:freetimeg} shows the ways participants spent their free time by gender. The top 3 activities among males and females were: i) listening to music or podcasts, ii) watching movies or TV shows, iii) spending time on social media platforms. Engaging in outdoor activities (e.g.~sports, hiking) was less common in females than males. Pursuing hobbies such as painting, cooking, and gardening was less common in both females and males. None of the females in the sample played video games during the free time. The majority of respondents reported multiple ways for spending their free time.

\begin{figure}
\includegraphics[width=1\linewidth]{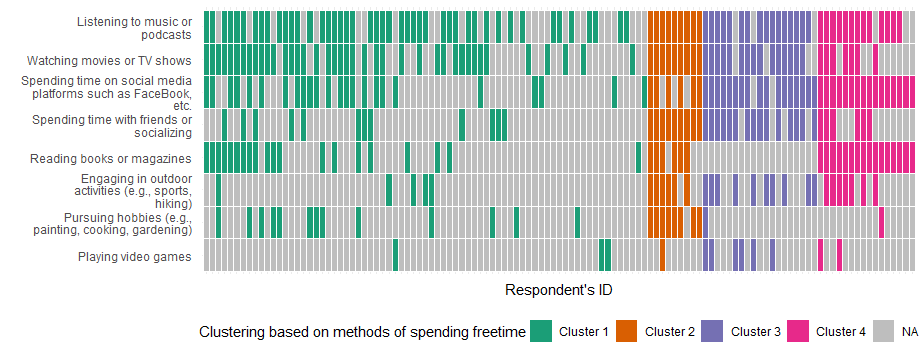} \caption{Heat map of responses to the multiple response question: How do you spend your free time? by clusters. The Y-axis represents the responses, while the X-axis represents the respondent IDs.  The clusters were mainly driven by the top five responses.}\label{fig:freetimec}
\end{figure}

Figure \ref{fig:freetimec} shows how the respondents were clustered according to their free time activities. According to Figure \ref{fig:freetimec} the clusters can be labeled as follows:

\textbf{Cluster 1: Audio-visual digital introverts} - High engagement in listening to music or podcast or watching movies or TV shows.

\textbf{Cluster 2: Audio-visual digital extroverts} - Everyone engages in activities such as listening to music or podcast, watching movies or TV shows, and spending time with friends.

\textbf{Cluster 3: Audio-visual digital extroverts but non book readers} - High engagement in activities such as listening to music or podcast, watching movies or TV shows, and spending time with friends, but neither engages in reading books and magazines.

\textbf{Cluster 4: Audio-visual digital extroverts but avid book readers} - Everyone spends time on social media platform and reading books.

\subsection{Stress management methods}\label{stress-management-methods-2}

\begin{figure}
\includegraphics[width=1\linewidth]{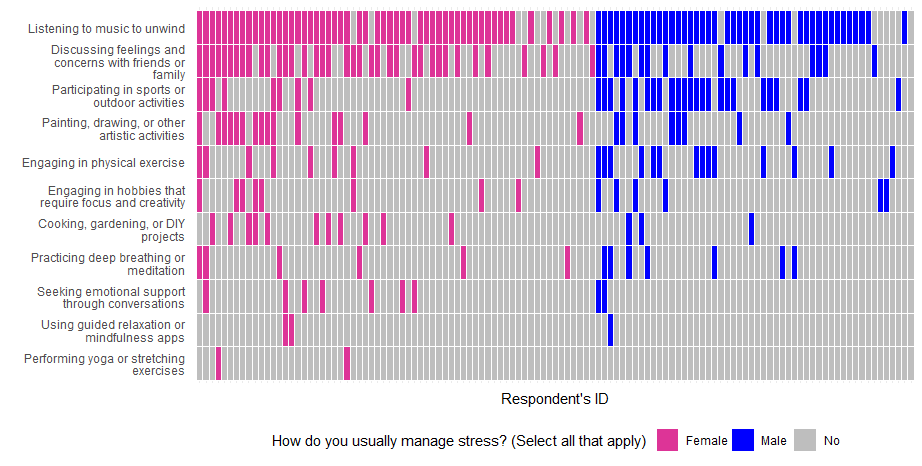} \caption{Heat map of responses to the multiple response question: How do you usually manage stress? by gender. The Y-axis represents the responses, while the X-axis represents the respondent IDs. Most males and females listen to music to unwind.}\label{fig:stressg}
\end{figure}

Figure \ref{fig:stressg} shows the ways of managing stress by gender. Listening music to unwind was the most common approach used by both male and female students. In addition to that, discussing feelings and concerns with friends or family was common among females, and participating in sports or outdoor activities was common in males.

\begin{figure}[H]
\includegraphics[width=1\linewidth]{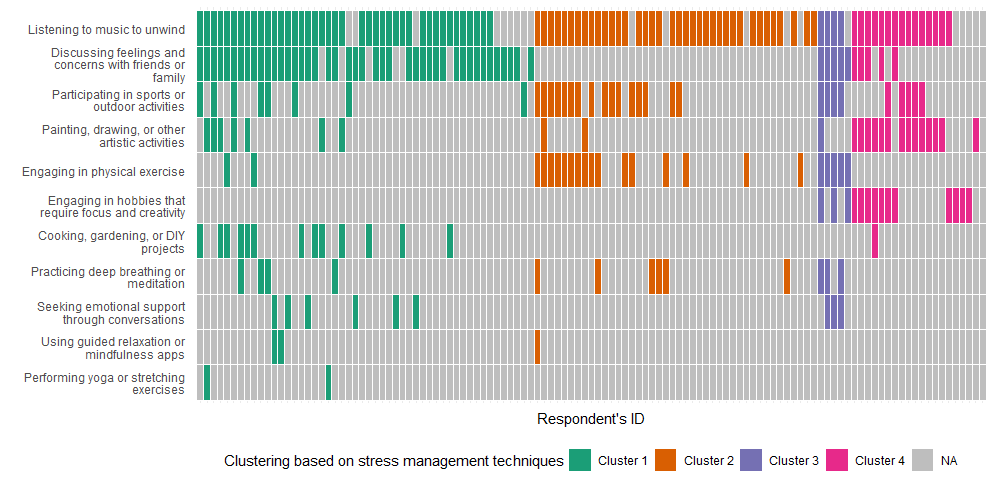} \caption{Heat map of responses to the multiple response question: How do you usually manage stress? by cluster. The Y-axis represents the responses, while the X-axis represents the respondent IDs. Listening to music and unwind was common for every cluster.}\label{fig:stressc}
\end{figure}

Figure \ref{fig:stressc} shows how individuals were clustered based on stress management methods. Four distinct clusters emerged. They can be interpreted as follows:

\textbf{Cluster 1: Social \& musical copers } - Mostly listening to music to unwind and discussing feelings and concerns with friends and family

\textbf{Cluster 2: Self-reliant copers} - Listening to music to unwind or engaging in physical exercise. They do not use the option ``Discussing feelings and concerns with friends and family'' to manage stress

\textbf{Cluster 3: Support-seekers copers} - Everyone uses the options discussing

\textbf{Cluster 4: Passive copers} - None of them engage in physical exercise, practice deep breathing or meditation, seek emotional support through conversations, use guided relaxation or mindfulness apps, perform yoga or stretching exercises.

\subsection{Studying techniques}\label{studying-techniques}

\begin{figure}[H]
\includegraphics[width=1\linewidth]{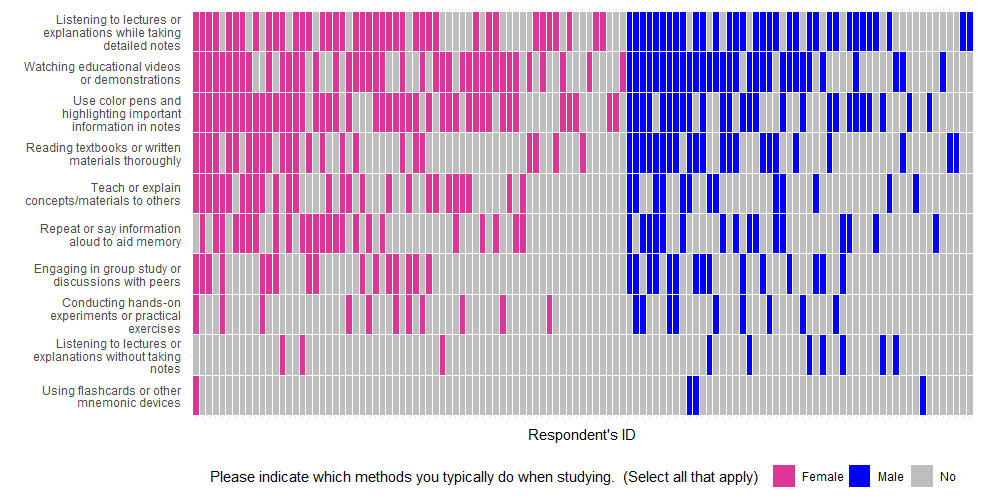} \caption{Heat map of responses to the multiple response question: 'Please indicate which methods you typically do when studying'  by gender. The Y-axis represents the responses, while the X-axis represents the respondent IDs.  Most males and females listen to lectures or explanations while taking detailed notes.}\label{fig:studyingg}
\end{figure}

According to Figure \ref{fig:studyingg}, both females and males reported at least one of the following study methods: (i) listening to lectures or explanations while taking detailed notes, (ii) watching educational videos or demonstrations, or (iii) using color pens and highlighting important information. This suggested that most individuals were read/write, auditory, and visual learners. There was no clear difference between the way females and males study.

\begin{figure}
\includegraphics[width=1\linewidth]{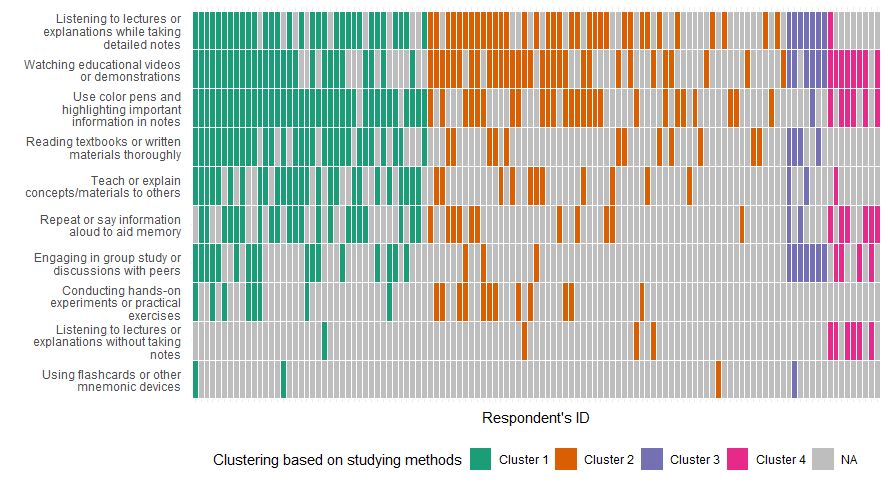} \caption{Heat map of responses to the multiple response question: Please indicate which methods you typically do when studying by cluster. The Y-axis represents the responses, while the X-axis represents the respondent IDs.}\label{fig:studyingc}
\end{figure}

According to Figure \ref{fig:studyingc}, the following features have been identified for the clusters based on their studying methods:

\textbf{Cluster 1: Read/write, visual and auditory learners} - They mostly listened to lectures or explanations while taking notes, watched educational videos and demonstrations, used colored pens and highlighted important information in their notes, and read textbooks or written materials thoroughly.

\textbf{Cluster 2: Visual and writing dominated learners} - They mostly listened to lecture/explanations while taking notes or watched educational videos and demonstrations. Further, they used colored pens, and highlighted important information in notes. These all involved a visual component. They rarely engaged in other components that did not have a writing component.

\textbf{Cluster 3: Auditory Writing dominated social learners} - All of them listened to lectures or explanations while taking detailed notes and engaged in group study or discussion with peers.

\textbf{Cluster 4: Auditory dominated learners} - Almost all of them watched educational videos and demonstrations, and furthermore, most of them do not take notes while listening to lectures or explanations. Furthermore, most of them repeated information aloud to aid memory.

\subsection{Methods of learning new things}\label{methods-of-learning-new-things-1}

\begin{figure}[H]
\includegraphics[width=1\linewidth]{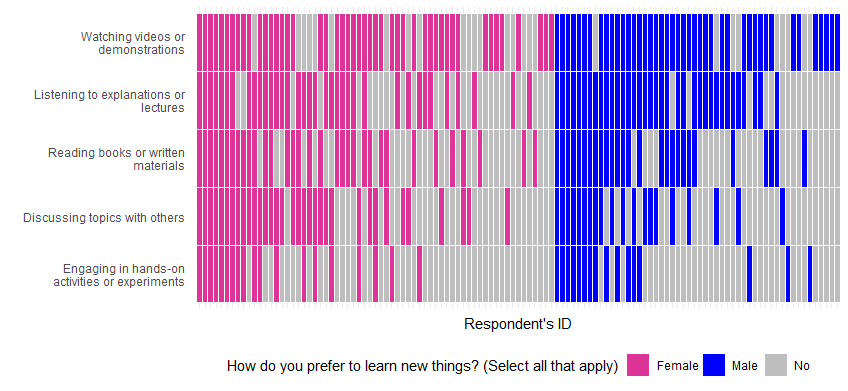} \caption{Heat map of responses to the multiple response question: How do you prefer to learn new things? by gender. The Y-axis represents the responses, while the X-axis represents the respondent IDs. Most males and females watch videos or demonstrations.}\label{fig:learningg}
\end{figure}

According to Figure \ref{fig:learningg}, watching videos and demonstrations was the most common method used by both males and females when learning new things. Females were more likely than males to discuss topics with others when learning new things.

\begin{figure}[H]
\includegraphics[width=1\linewidth]{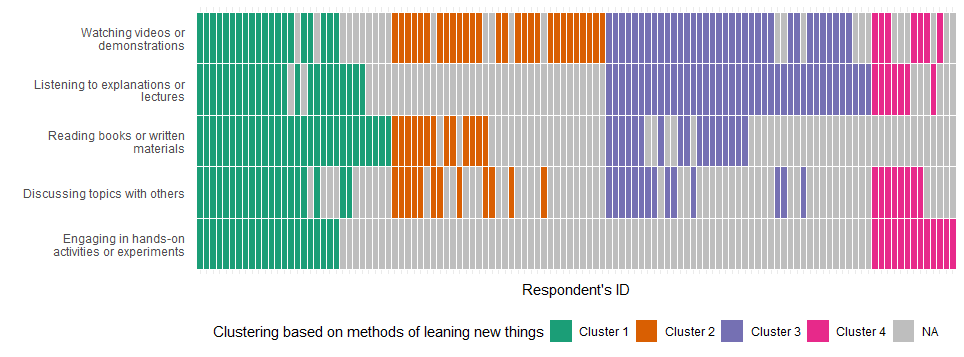} \caption{Heat map of responses to the multiple response question: How do you prefer to learn new things? by cluster. The Y-axis represents the responses, while the X-axis represents the respondent IDs. Most males and females watch videos or demonstrations.}\label{fig:learningc}
\end{figure}

According to Figure \ref{fig:learningc} there can be seen clear distinction between the four clusters.

\textbf{Cluster 1: Read/write dominated learners} - All of them use reading books or written materials for leaning new things

\textbf{Cluster 2: Visual dominated learners} - They either use watching videos or demonstrations or discussing topics with other. They do not listen to explanations or lectures and engage in hand-on activities and experiments.

\textbf{Cluster 3: Auditory dominated learners} - All of them listening to explanations or lectures. They do not listen to explanations or lectures and engage in hand-on activities and experiments or read books or written materials.

\textbf{Cluster 4: Kinesthetic dominated learners} - All of them engage in hands-on activities or experiments. They do not engage in hand-on activities and experiments.

\subsection{Results of relationships between studying methods, methods of learning new things, leisure activities and stress management methods}\label{results-of-relationships-between-studying-methods-methods-of-learning-new-things-leisure-activities-and-stress-management-methods}

\begin{figure}
\includegraphics[width=1\linewidth]{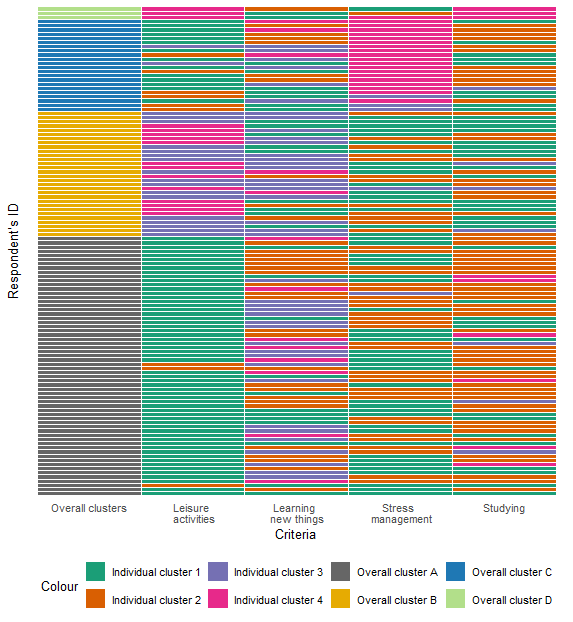} \caption{Comparison of overall clusterings and individual culsterings. Overall clustering is based on the results of individual clusters from each dimension. The Y-axis shows both the overall cluster assignments and the clusters from individual dimensions, while the X-axis represents respondent IDs.}\label{fig:all}
\end{figure}

Figure \ref{fig:all} compares cluster memberships across four dimensions; i) studying, ii) learning new things, iii) leisure activities, and iv) stress management techniques, along with the overall clustering derived from these individual results. The plot reveals a clear lack of consistency across the domains. For instance, individuals who were classified into one cluster based on their methods of learning new things were not grouped in the same cluster for studying, stress management, or leisure activities. This means that individuals who shared similar characteristics or behaviors in one dimension (e.g., methods of learning new things) were not necessarily aligned with the same group when it comes to other dimensions, such as studying, stress management, or leisure time activities. For example, according to Figure \ref{fig:all}, individuals who were similar in terms of leisure time activities were quite different when it comes to methods used to learn new things. This indicates the variability in how individuals approach different aspects of their lives. This meant that individuals who were similar in one dimension did not necessarily belong to the same cluster in other dimensions.

\section{Discussion}\label{discussion}

This research aims to identify the most commonly preferred leisure time activities, stress management methods, studying Methods, methods of learning new things among statistics students, while examining any notable differences between male and female preferences. Furthermore, we explore the relationships between these dimensions to understated the interplay between these components.

\subsection{The most commonly preferred leisure time activities, stress management methods, studying Methods, methods of learning new things among statistics students, and differences between male and female preferences}\label{the-most-commonly-preferred-leisure-time-activities-stress-management-methods-studying-methods-methods-of-learning-new-things-among-statistics-students-and-differences-between-male-and-female-preferences}

The most common leisure activities among both females and males are listening to music or podcasts, watching movies or TV shows, and spending time on social media platforms such as Facebook, etc. The results show males and females predominantly engage in passive forms of activities. This highlights the widespread influence of digital media among youngsters. Furthermore, pursuing creative or skill-based hobbies such as painting, cooking, and gardening are relatively uncommon among both males and females. This suggests that free time is more frequently spent on media consumption rather than activities that require mental or physical effort and skill development. Furthermore, we observe females are more likely to engage in passive leisure activities than males. This is consistent with the results of Kim, Brown, and Yang (2019). Terzi et al. (2024) also reported significant relationships between gender, free time management, and life quality.

The most common coping strategy used by both male and female students is listening to music. Discussing feelings and concerns with family and friends is the second most common stress coping method among undergraduates. This aligns with Ahmad, Md Yusoff, and Razak (2011) who found that talking to friends was a leading stress reduction technique for students across all academic years. Notably, this study was also conducted in South Asia, where cultural similarities may contribute to the prevalence of this coping strategy. This contradicts to studies done on medical, nursing, and dentistry fields. Sikander and Aziz (2012) also reported that the most common coping strategy was discuss feeling with friends or class mates. In those health science fields the most common methods include seeking help and emotion-focused strategies (Vally et al. 2018).

Listening to lectures or explanations while taking notes, watching educational videos and demonstrations, and using colors to highlight important notes are the most commonly used methods in studying.

Watching videos and demonstration is the most preferred method of learning new things among both females and males. When learning new things and studying, most of the students use multiple modes. This is inline with the results of Lujan and DiCarlo (2006). In our analysis students mostly use passive modes for learning; engagement in kinesthetic activities and social interaction activities, such as discussions with friends, is very low. This contradicts the findings of Yousef (2016), in which their sample showed a balanced preference for the four dimensions of learning styles: active/reflective, sensing/intuitive, sequential/global.

Across all four dimensions we examined, the top activities are either listening to music or lectures and watching videos or TV shows, suggesting that individuals are introverted and passive learners.

\subsection{Relationship between leisure time activities, stress management methods, studying Methods, and methods of learning new things}\label{relationship-between-leisure-time-activities-stress-management-methods-studying-methods-and-methods-of-learning-new-things}

We did not find a relationship between Leisure time activities, stress management methods, studying Methods, and methods of learning new things. These individuals tend to use different approaches across these dimensions. For example, someone who belongs to a specific cluster based on their preferred method of learning new things may not fall into the same cluster when it comes to studying, managing stress, or spending leisure time. This suggests that individuals who share similar behaviors or preferences in one area do not necessarily exhibit the same patterns in others. Lehto et al. (2014) also stated that there seems to be
no direct causal relationship between student stress and their leisure participation.

\subsection{Implications for educational research and practices}\label{implications-for-educational-research-and-practices}

The lack of a strong relationship between the leisure activities, stress management methods, study methods, and methods of learning new things suggests that individual adaption and coping strategies are independent and unique to different situations. Hence, support programs for these four dimensions should be multifaceted and personalized rather than relying on a one-size-fit grouping for all dimensions. By recognizing these clusters and relationships, educators can create inclusive learning environments that enhance student engagement and comprehension. Furthermore, when teaching subjects like statistics, instructors should carefully design lesson plans that foster collaboration and promote kinesthetic learning, which can help students build essential social and soft skills. This ensures that students are better prepared and ready to engage with society. In addition, awareness programs should be established to encourage physical exercise, thereby supporting students' overall well-being.

\subsection{Limitations}\label{limitations}

One of the main limitation of the study is, the study only considered age and gender as the demographic factors. In addition to that other factors such as district, number of siblings, travelling from home or stay in hostel could be included for further analysis. Furthermore, this research was a cross-sectional study conducted at a specific point in time. This design may limit the ability to determine variability of these four dimensions over time or causality in the correlation. For future research, longitudinal or experimental designs could provide more robust results for subsequent studies. To further understand the students preferences instead of asking students mark yes/no setting on different options, responses could be based on 1-5 scale with 1 being `Never' and 5 for frequently.

\section{Conclusion}\label{conclusions}

Overall, the results reflect a strong preference for digital and passive activities across all the four dimensions (studying methods, methods
of ;earning new things, leisure activities,
and stress management methods) among both males and females. The absence of strong relationships among students' learning styles, stress management methods, leisure activities, and approaches to learning new things suggests that that these behaviors are highly individualized rather than aligned consistent profiles. Educational or support interventions should therefore avoid one-size-fits-all strategies and instead adopt more personalized or flexible approaches.

The results shed light on the field of education by offering insights into student behaviors on leisure activities, stress management methods, studying and methods of learning new things. This study also highlights how statistical visualizations can aid in understanding and communicating these individual differences effectively.

\section*{Transparency and openness}\label{transparency-and-openness}
\addcontentsline{toc}{section}{Transparency and openness}

The data reproduce the results are package in to a R (R Core Team (2024a)) software package called \texttt{LearnLifeStat} and available at \url{https://github.com/thiyangt/LearnLifeStat}.

\section*{Declaration of generative AI in scientific writing}\label{declaration-of-generative-ai-in-scientific-writing}
\addcontentsline{toc}{section}{Declaration of generative AI in scientific writing}

During the preparation of this work the author used ChatGPT in order to improve readability and language. After using this tool/service, the author reviewed and edited the content as needed and take full responsibility for the content of the published article.

\section*{References}\label{references}
\addcontentsline{toc}{section}{References}

\phantomsection\label{refs}


\begin{CSLReferences}{1}{0}
\bibitem[\citeproctext]{ref-abdelhafez2021learning}
Abdelhafez, Amal Ismael, Nermine M Elcokany, Asmaa Saber Ghaly, and Fahima Akhter. 2021. {``Learning Environment Stressors and Coping Styles Among Nursing Students.''} \emph{International Journal of Nursing Education} 12 (4): 6--14.

\bibitem[\citeproctext]{ref-adasi2020gender}
Adasi, Grace S, Kwaku D Amponsah, Salifu M Mohammed, Rita Yeboah, and Priscilla C Mintah. 2020. {``Gender Differences in Stressors and Coping Strategies Among Teacher Education Students at University of Ghana.''} \emph{Journal of Education and Learning} 9 (2): 123--33.

\bibitem[\citeproctext]{ref-agahi2008leisure}
Agahi, Neda, and Marti G Parker. 2008. {``Leisure Activities and Mortality: Does Gender Matter?''} \emph{Journal of Aging and Health} 20 (7): 855--71.

\bibitem[\citeproctext]{ref-ahmad2011stress}
Ahmad, Mas Suryalis, MM Md Yusoff, and I Abdul Razak. 2011. {``Stress and Its Relief Among Undergraduate Dental Students in Malaysia.''} \emph{Southeast Asian Journal of Tropical Medicineand Public Health} 42 (4): 996.

\bibitem[\citeproctext]{ref-al2018stress}
Al-Gamal, Ekhlas, Aisha Alhosain, and Khulood Alsunaye. 2018. {``Stress and Coping Strategies Among Saudi Nursing Students During Clinical Education.''} \emph{Perspectives in Psychiatric Care} 54 (2): 198--205.

\bibitem[\citeproctext]{ref-alshahrani2018undergraduate}
Alshahrani, Yousef, Lynette Cusack, and Philippa Rasmussen. 2018. {``Undergraduate Nursing Students' Strategies for Coping with Their First Clinical Placement: Descriptive Survey Study.''} \emph{Nurse Education Today} 69: 104--8.

\bibitem[\citeproctext]{ref-amponsah2020stressors}
Amponsah, Kwaku Darko, Grace Sintim Adasi, Salifu Maigari Mohammed, Ernest Ampadu, and Abraham Kwadwo Okrah. 2020. {``Stressors and Coping Strategies: The Case of Teacher Education Students at University of Ghana.''} \emph{Cogent Education} 7 (1): 1727666.

\bibitem[\citeproctext]{ref-athenstaedt2009gender}
Athenstaedt, Ursula, Gerold Mikula, and Cornelia Bredt. 2009. {``Gender Role Self-Concept and Leisure Activities of Adolescents.''} \emph{Sex Roles} 60: 399--409.

\bibitem[\citeproctext]{ref-avital2017gender}
Avital, Dana. 2017. {``Gender Differences in Leisure Patterns at Age 50 and Above: Micro and Macro Aspects.''} \emph{Ageing \& Society} 37 (1): 139--66.

\bibitem[\citeproctext]{ref-bistricky2018understanding}
Bistricky, Steven L, Kristina L Harper, Caroline M Roberts, Diana M Cook, Staci L Schield, Jennifer Bui, and Mary B Short. 2018. {``Understanding and Promoting Stress Management Practices Among College Students Through an Integrated Health Behavior Model.''} \emph{American Journal of Health Education} 49 (1): 12--27.

\bibitem[\citeproctext]{ref-caldwell1992relationship}
Caldwell, Linda L, Edward A Smith, and Ellen Weissinger. 1992. {``The Relationship of Leisure Activities and Perceived Health of College Students.''} \emph{Loisir Et Soci{é}t{é}/Society and Leisure} 15 (2): 545--56.

\bibitem[\citeproctext]{ref-chan2023college}
Chan, Wing Yi, Anthony Rodriguez, Regina A Shih, Joan S Tucker, Eric R Pedersen, Rachana Seelam, and Elizabeth J D'Amico. 2023. {``How Do College Students Use Their Free Time? A Latent Profile Analysis of Leisure Activities and Substance Use.''} \emph{Leisure Sciences} 45 (4): 331--50.

\bibitem[\citeproctext]{ref-dominguez2025data}
Domı́nguez, Lauren Genith Isaza, Antonio Robles-Gómez, and Rafael Pastor-Vargas. 2025. {``A Data-Driven Approach to Engineering Instruction: Exploring Learning Styles, Study Habits, and Machine Learning.''} \emph{IEEE Access}.

\bibitem[\citeproctext]{ref-dunn1990understanding}
Dunn, Rita. 1990. {``Understanding the Dunn and Dunn Learning Styles Model and the Need for Individual Diagnosis and Prescription.''} \emph{Reading, Writing, and Learning Disabilities} 6 (3): 223--47.

\bibitem[\citeproctext]{ref-espinoza2019vark}
Espinoza-Poves, Jenny L, Walter A Miranda-Vı́lchez, and Raquel Chafloque-Céspedes. 2019. {``The Vark Learning Styles Among University Students of Business Schools.''} \emph{Journal of Educational Psychology-Propositos y Representaciones} 7 (2): 401--15.

\bibitem[\citeproctext]{ref-felder1988}
Felder, R M, and L K Silverman. n.d. {``Learning and Teaching Styles in Engineering Education.''} \emph{Engineering Education} 78 (7): 674--81.

\bibitem[\citeproctext]{ref-felder1993reaching}
Felder, Richard M. 1993. {``Reaching the Second Tier.''} \emph{Journal of College Science Teaching} 23 (5): 286--90.

\bibitem[\citeproctext]{ref-felder2002learning}
---------. 2002. {``Learning and Teaching Styles in Engineering Education.''}

\bibitem[\citeproctext]{ref-finkel2018gender}
Finkel, Deborah, Ross Andel, and Nancy L Pedersen. 2018. {``Gender Differences in Longitudinal Trajectories of Change in Physical, Social, and Cognitive/Sedentary Leisure Activities.''} \emph{The Journals of Gerontology: Series B} 73 (8): 1491--1500.

\bibitem[\citeproctext]{ref-fleming2006vark}
Fleming, Neil D. 2006. {``VARK Visual, Aural/Auditory, Read/Write, Kinesthetic.''} \emph{New Zealand: Bonwell Green Mountain Falls}.

\bibitem[\citeproctext]{ref-fosnacht2018first}
Fosnacht, Kevin, Alexander C McCormick, and Rosemarie Lerma. 2018. {``First-Year Students' Time Use in College: A Latent Profile Analysis.''} \emph{Research in Higher Education} 59 (7): 958--78.

\bibitem[\citeproctext]{ref-gougucs2020scoring}
Göğüş, Aytaç, and Gürdal Ertek. 2020. {``A Scoring Approach for the Assessment of Study Skills and Learning Styles.''} \emph{International Journal of Information and Education Technology} 10 (10): 715--22.

\bibitem[\citeproctext]{ref-hendel2004undergraduate}
Hendel, Darwin D, and Roger D Harrold. 2004. {``Undergraduate Student Leisure Interests over Three Decades.''} \emph{College Student Journal} 38 (4): 557--69.

\bibitem[\citeproctext]{ref-honey1989learning}
Honey, Peter, and Alan Mumford. 1989. \emph{Learning Styles Questionnaire}. Organization Design; Development, Incorporated.

\bibitem[\citeproctext]{ref-hulteen2017global}
Hulteen, Ryan M, Jordan J Smith, Philip J Morgan, Lisa M Barnett, Pedro C Hallal, Kim Colyvas, and David R Lubans. 2017. {``Global Participation in Sport and Leisure-Time Physical Activities: A Systematic Review and Meta-Analysis.''} \emph{Preventive Medicine} 95: 14--25.

\bibitem[\citeproctext]{ref-idrizi2023gender}
Idrizi, Ermira, Sonja Filiposka, and Vladimir Trajkovikj. 2023. {``Gender Impact on STEM Online Learning-a Correlational Study of Gender, Personality Traits and Learning Styles in Relation to Different Online Teaching Modalities.''} \emph{Multimedia Tools and Applications} 82 (19): 30201--19.

\bibitem[\citeproctext]{ref-jaju2002learning}
Jaju, Anupam, Hyokjin Kwak, and George M Zinkhan. 2002. {``Learning Styles of Undergraduate Business Students: A Cross-Cultural Comparison Between the US, India, and Korea.''} \emph{Marketing Education Review} 12 (2): 49--60.

\bibitem[\citeproctext]{ref-katsioloudis2012comparative}
Katsioloudis, Petros, and Todd D Fantz. 2012. {``A Comparative Analysis of Preferred Learning and Teaching Styles for Engineering, Industrial, and Technology Education Students and Faculty.''} \emph{Journal of Technology Education} 23 (2).

\bibitem[\citeproctext]{ref-kim2018associations}
Kim, Jong-Ho, and Stephen L Brown. 2018. {``The Associations Between Leisure, Stress, and Health Behavior Among University Students.''} \emph{American Journal of Health Education} 49 (6): 375--83.

\bibitem[\citeproctext]{ref-kim2019types}
Kim, Jong-Ho, Stephen L Brown, and Heewon Yang. 2019. {``Types of Leisure, Leisure Motivation, and Well-Being in University Students.''} \emph{World Leisure Journal} 61 (1): 43--57.

\bibitem[\citeproctext]{ref-klement2014my}
Klement, Milan. 2014. {``How Do My Students Study? An Analysis of Students' of Educational Disciplines Favorite Learning Styles According to VARK Classification.''} \emph{Procedia-Social and Behavioral Sciences} 132: 384--90.

\bibitem[\citeproctext]{ref-kolb1976management}
Kolb, David A. 1976. {``Management and the Learning Process.''} \emph{California Management Review} 18 (3): 21--31.

\bibitem[\citeproctext]{ref-kuo2014relationships}
Kuo, Tingya, and Hung-Lian Tang. 2014. {``Relationships Among Personality Traits, Facebook Usages, and Leisure Activities--a Case of Taiwanese College Students.''} \emph{Computers in Human Behavior} 31: 13--19.

\bibitem[\citeproctext]{ref-lehto2014student}
Lehto, Xinran Y, Ounjoung Park, Xiaoxiao Fu, and Gyehee Lee. 2014. {``Student Life Stress and Leisure Participation.''} \emph{Annals of Leisure Research} 17 (2): 200--217.

\bibitem[\citeproctext]{ref-lujan2006first}
Lujan, Heidi L, and Stephen E DiCarlo. 2006. {``First-Year Medical Students Prefer Multiple Learning Styles.''} \emph{Advances in Physiology Education} 30 (1): 13--16.

\bibitem[\citeproctext]{ref-cluster}
Maechler, Martin, Peter Rousseeuw, Anja Struyf, Mia Hubert, and Kurt Hornik. 2023. \emph{Cluster: Cluster Analysis Basics and Extensions}. \url{https://CRAN.R-project.org/package=cluster}.

\bibitem[\citeproctext]{ref-mavsic2020relationship}
Mašic, Adela, Edda Polz, and Senad Becirovic. 2020. {``The Relationship Between Learning Styles, GPA, School Level and Gender.''} \emph{Online Submission} 11 (1): 51--60.

\bibitem[\citeproctext]{ref-moayyeri2015impact}
Moayyeri, Hessam. 2015. {``The Impact of Undergraduate Students' Learning Preferences (VARK Model) on Their Language Achievement.''} \emph{Journal of Language Teaching \& Research} 6 (1).

\bibitem[\citeproctext]{ref-mosley1994stress}
Mosley Jr, Thomas H, Sean G Perrin, Susan M Neral, Patricia M Dubbert, Carol A Grothues, and Bernadine M Pinto. 1994. {``Stress, Coping, and Well-Being Among Third-Year Medical Students.''} \emph{Academic Medicine} 69 (9): 765--67.

\bibitem[\citeproctext]{ref-naik2013influence}
Naik, Bijayananda. 2013. {``Influence of Culture on Learning Styles of Business Students.''} \emph{International Journal of Education Research} 8 (1).

\bibitem[\citeproctext]{ref-othman2010different}
Othman, Norasmah, and Mohd Hasril Amiruddin. 2010. {``Different Perspectives of Learning Styles from VARK Model.''} \emph{Procedia-Social and Behavioral Sciences} 7: 652--60.

\bibitem[\citeproctext]{ref-pashler2008learning}
Pashler, Harold, Mark McDaniel, Doug Rohrer, and Robert Bjork. 2008. {``Learning Styles: Concepts and Evidence.''} \emph{Psychological Science in the Public Interest} 9 (3): 105--19.

\bibitem[\citeproctext]{ref-prithishkumar2014understanding}
Prithishkumar, Ivan J, and Stelin Agnes Michael. 2014. {``Understanding Your Student: Using the VARK Model.''} \emph{Journal of Postgraduate Medicine} 60 (2).

\bibitem[\citeproctext]{ref-r}
R Core Team. 2024a. \emph{R: A Language and Environment for Statistical Computing}. Vienna, Austria: R Foundation for Statistical Computing. \url{https://www.R-project.org/}.

\bibitem[\citeproctext]{ref-ggplot2}
---------. 2024b. \emph{R: A Language and Environment for Statistical Computing}. Vienna, Austria: R Foundation for Statistical Computing. \url{https://www.R-project.org/}.

\bibitem[\citeproctext]{ref-ramos2011comparison}
Ramos, Jose A. 2011. {``A Comparison of Perceived Stress Levels and Coping Styles of Non-Traditional Graduate Students in Distance Learning Versus on-Campus Programs.''} \emph{Contemporary Educational Technology} 2 (4): 282--93.

\bibitem[\citeproctext]{ref-rohani2024identifying}
Rohani, Narjes, Stephen Sowa, Areti Manataki, et al. 2024. {``Identifying Learning Preferences and Strategies in Health Data Science Courses: Systematic Review.''} \emph{JMIR Medical Education} 10 (1): e50667.

\bibitem[\citeproctext]{ref-seemiller2019generation}
Seemiller, Corey, Meghan Grace, Paula Dal Bo Campagnolo, Isa Mara Da Rosa Alves, and Gustavo Severo De Borba. 2019. {``How Generation z College Students Prefer to Learn: A Comparison of US and Brazil Students.''} \emph{Journal of Educational Research and Practice} 9 (1): 25.

\bibitem[\citeproctext]{ref-sikander2012stressors}
Sikander, Shomail, and Faisal Aziz. 2012. {``Stressors and Coping Strategies Among Baccalaureate Nursing Students at Shifa College of Nursing Islamabad, Pakistan.''} \emph{International Journal of Nursing Education} 4 (2).

\bibitem[\citeproctext]{ref-skinner2003searching}
Skinner, Ellen A, Kathleen Edge, Jeffrey Altman, and Hayley Sherwood. 2003. {``Searching for the Structure of Coping: A Review and Critique of Category Systems for Classifying Ways of Coping.''} \emph{Psychological Bulletin} 129 (2): 216.

\bibitem[\citeproctext]{ref-terzi2024university}
Terzi, Esranur, Utku Isik, Berat Can Inan, Can Akyildiz, and Umit Dogan Ustun. 2024. {``University Students' Free Time Management and Quality of Life: The Mediating Role of Leisure Satisfaction.''} \emph{BMC Psychology} 12 (1): 239.

\bibitem[\citeproctext]{ref-vally2018comparative}
Vally, Zahir, Brettjet L Cody, Safeya NM Alsheraifi, and Maryam A Albloshi. 2018. {``A Comparative Description of Perceived Stress and Coping Strategies Among Psychology and Nonpsychology Students in the United Arab Emirates.''} \emph{Perspectives in Psychiatric Care} 54 (4): 539--46.

\bibitem[\citeproctext]{ref-van2000felder}
Van Zwanenberg, Nigel, LJ Wilkinson, and Amy Anderson. 2000. {``Felder and Silverman's Index of Learning Styles and Honey and Mumford's Learning Styles Questionnaire: How Do They Compare and Do They Predict Academic Performance?''} \emph{Educational Psychology} 20 (3): 365--80.

\bibitem[\citeproctext]{ref-waterhouse2024university}
Waterhouse, Philippa, and Rajvinder Samra. 2024. {``University Students' Coping Strategies to Manage Stress: A Scoping Review.''} \emph{Educational Review}, 1--41.

\bibitem[\citeproctext]{ref-wehrwein2007gender}
Wehrwein, Erica A, Heidi L Lujan, and Stephen E DiCarlo. 2007. {``Gender Differences in Learning Style Preferences Among Undergraduate Physiology Students.''} \emph{Advances in Physiology Education} 31 (2): 153--57.

\bibitem[\citeproctext]{ref-tidyverse}
Wickham, Hadley, Mara Averick, Jennifer Bryan, Winston Chang, Lucy D'Agostino McGowan, Romain François, Garrett Grolemund, et al. 2019. {``Welcome to the {tidyverse}.''} \emph{Journal of Open Source Software} 4 (43): 1686. \url{https://doi.org/10.21105/joss.01686}.

\bibitem[\citeproctext]{ref-yousef2016learning}
Yousef, Darwish Abdulrahman. 2016. {``Learning Styles Preferences of Statistics Students: A Study in the Faculty of Business and Economics at the UAE University.''} \emph{Quality Assurance in Education} 24 (2): 227--43.

\end{CSLReferences}
\end{document}